\documentclass[pra,twocolumn,showpacs,preprintnumbers,superscriptaddress]{revtex4}
\usepackage{bm}
\usepackage{graphicx}
\usepackage{amsbsy}
\usepackage{amsmath}
\usepackage{amsfonts}
\usepackage{amsthm}

\date{\today}

\begin{document}

\theoremstyle{plain}
\newtheorem{theorem}{Theorem}
\newtheorem{lemma}[theorem]{Lemma}
\newtheorem{corollary}[theorem]{Corollary}
\newtheorem{proposition}[theorem]{Proposition}
\newtheorem{conjecture}[theorem]{Conjecture}

\theoremstyle{definition}
\newtheorem{definition}{Definition}

\theoremstyle{remark}
\newtheorem*{remark}{Remark}
\newtheorem{example}{Example}

\title{Highly nonclassical photon statistics in parametric down conversion}

\author{Edo Waks}
\affiliation{E. L. Ginzton Laboratories, Stanford University, Stanford, California 94305, USA}
\author{Barry C.\ Sanders}
\affiliation{Institute for Quantum Information Science, University of Calgary, Alberta T2N 1N4, Canada}
\affiliation{Centre for Quantum Computer Technology, Macquarie University, New South Wales 2109,
    Australia}
\author{Eleni Diamanti}
\affiliation{E. L. Ginzton Laboratories, Stanford University, Stanford, California 94305, USA}
\author{Yoshihisa Yamamoto}
\affiliation{E. L. Ginzton Laboratories, Stanford University, Stanford, California 94305, USA}

\begin{abstract}
We use photon counters to obtain the joint photon counting statistics from
twin-beam non-degenerate parametric down conversion, and we demonstrate directly,
and with no auxiliary assumptions, that these twin beams are nonclassical.
\end{abstract}

\maketitle

\section{introduction}

The quantum nature of light is exemplified by the fact that it can
exhibit both undular (wavelike) and corpuscular (particlelike)
properties under different circumstances. Using an atomic cascade
to produce a pair of nearly simultaneous photons~\cite{Koc67},
Grangier et al~\cite{Gra86} demonstrated that a heralded photon
can either interfere with itself in a Mach-Zehnder interferometer
or be shown to either wholly transmit or reflect by a beam
splitter, which demonstrates complementarity for light. More
recently the continuous trade-off of undular and corpuscular
features in a quantum non-demolition measurement~\cite{San89} was
demonstrated using a non-deterministic linear optical
gate~\cite{Pry04}. Technically, nonclassical light corresponds to
field states whose Glauber-Sudarshan $P$-representation is
singular~\cite{Sud63,Gla63}; practically nonclassicality can be
established by demonstrating photon antibunching~\cite{Kim77} or
using optical homodyne tomography~\cite{Leo97,Lvo05} to
demonstrate that the Wigner function (which is a Gaussian
convolution of the $P$-representation) exhibits
squeezing~\cite{Smi93} or is negative over some region~\cite{Lvo01}.

Recently Waks, Diamanti, Sanders, Bartlett, and Yamamoto
(WDSBY)~\cite{Wak04} exploited cutting edge photon counting
technology (using the Visible Light Photon Counter~\cite{Wak03})
to demonstrate that a single beam emitted by parametric down
conversion (PDC) is nonclassical by measuring its photon
statistics and avoiding auxiliary assumptions such as detector
efficiency. Here we reinterpret the WDSBY data using Klyshko's
powerful criterion for direct detection of nonclassicality via
photon statistics~\cite{Kly96}, and we demonstrate nonclassicality
of twin beam PDC by using two photon counters and Lee's extension
of Klyshko's criterion to two beams~\cite{Lee98}.

Although the simultaneity of photon pair production by twin-beam,
or nondegenerate, PDC is well known~\cite{Bur70}, leading to its
applications as a heralded photon source~\cite{Hon86} and for
tests of Bell's inequalities~\cite{Ou88}, the joint photon
statistics of the twin beams have not previously been directly
measured. These statistics are necessary for a direct confirmation
of the nonclassicality of twin-beam PDC, which is one of the most
important tools in quantum optics as a source of correlated or
entangled pairs of photons~\cite{Har96} and, more recently, as a
source of entangled four-tuples of photons~\cite{Wal04}. We have
performed this direct photon counting experiment and obtained
joint photon counting data for the twin beams and thereby showed
that these twin beams are truly nonclassical.

\section{Single beam non-classical statistics}

Theoretical predictions suggest that the beams should be
nonclassical, but theories generally treat the twin-beam PDC
output as a two-mode squeezed vacuum state~\cite{Sch85}
\begin{equation}
\label{eq:2modesqvac}
    |\eta\rangle=\sqrt{1-|\eta|^2}\sum_{k=0}^\infty \eta^k|kk\rangle,
\end{equation}
with $\eta\in\mathbb{C}$ and the argument of $\eta$ dependent on
the phase of the pump laser field for the PDC.  For degenerate
PDC, the state~(\ref{eq:2modesqvac}) is replaced by
\begin{equation}
    |\chi\rangle=\sqrt{1-|\chi|^2}\sum_{k=0}^\infty\chi^k|2k\rangle,
\end{equation}
and the photon number distribution is
\begin{equation}
    p_n=(1-|\chi|^2)|\chi|^{2n}.
\end{equation}
This state corresponds to a joint
photon number distribution for the two beams, which is given by
\begin{equation}
\label{eq:2modecounting}
    p_{n_1,n_2}=(1-|\eta|^2)|\eta|^{2n_1}\delta_{n_1,n_2}.
\end{equation}
Klyshko considered a state of the type emitted by an ideal
degenerate PDC, and Lee's analysis of photon counting correlations
explicitly assumed two-mode states and included a treatment of the
two-mode squeezed vacuum~(\ref{eq:2modesqvac}).

The reality of PDC and photon counting is somewhat more
complicated.  The twin beams (known as signal and idler fields)
are each multimode so the theoretical treatments of these outputs
are not immediately applicable. In fact there are so many modes
for each of the signal and idler beams that we observe a photon
distribution over these modes that is Poissonian. The Poisson
distribution originates from the fact that there so many signal
and idler modes that the average number of pairs per mode is much
less than one. This means that each mode is predominantly a vacuum
state with a small one-photon contribution and a negligible
multi-photon contribution.  Thus, the generated signal is created
by a sum of independent spontaneous emitters (one for each mode),
and, as each pair is created independently, the pair creation
statistics must be a Poisson distribution.

This discrepancy in photon statistics between the usual theoretical treatment of PDC
output~\cite{Sch85} and experimental reality makes the direct
measurement of joint photon counting statistics and verification
of nonclassicality even more interesting and important. Properties
of twin-beam PDC have been explored in an elegant experiment by
Haderka et al~\cite{Had05}, but their approach is different in
that they do not use two distinct photon counters on each beam.
Thus, inferences of nonclassicality rely on auxiliary assumptions
that are not required by our approach such as taking into account the
losses during transmission, quantum efficiency and internal noise
of the camera and noise due to other light sources to obtain the joint
signal-idler photon number distribution~\cite{Had05}.

To begin, we re-analyze the WDSBY data for single beam PDC.
WDSBY~\cite{Wak04} conclusively demonstrated nonclassicality using
their data for one-, two-, and three-photon measured
counts~$\{\wp_n;n=1,2,3\}$, where we use the symbol~$p$ for the
ideal probability and~$\wp$ for the measured probability. We will
now compare their results to Klyshko's criterion~\cite{Kly96}. We
introduce Klyshko's criterion for two reasons:
(i)~Klyshko's
criterion demonstrates that the WDSBY data violates classical
bounds beyond the three-photon case studied by WDSBY, and
(ii)~Klyshko's criterion is a basis for Lee's
criterion~\cite{Lee98} which we use for the analysis of the twin
beams.

WDSBY used the fact that nonclassicality of light, defined by the
singularity of the Glauber-Sudarshan $P$-representation holds if
and only if the photocount distribution for the beam cannot be
expressed as a sum or integral of Poisson distributions (which
corresponds to classical count distributions). Specifically the
photon count statistics~$\{\wp_n;n=1,2,3\}$ were irreconcilable
with classical photon statistics; thus photon counting provided a
direct means for establishing that a single-beam light source is
nonclassical.

To demonstrate nonclassical photon statistics, WDSBY pumped a
Type-I phase-matched BBO crystal set up for collinear degenerate
amplification with 20ns pulses of the fourth harmonic (266nm) of a
Q-switched Nd:YAG laser.  In this configuration, the
down-converted photons had half the energy of the pump (532nm) and
travelled in the same direction. The pump was removed by a prism,
while the down-conversion was focused by a 250mm lens onto the
VLPC detector.  The detector output was amplified and then sent to
a gated boxcar integrator, which was triggered by the laser.  This
configuration was used to measure the photon number distribution
of the detected field. WDSBY achieved a forty standard deviation
violation of classicality, according to the criterion
\begin{equation}
\label{eq:GammaWDSBY}
    \Gamma=\frac{\wp_2}{\wp_1+\wp_2+\wp_3}>\Gamma_\text{classical}
        \equiv\frac{3}{3+2\sqrt{6}}\approx 0.379
\end{equation}
without any auxiliary assumptions about the detector or field.

In fact Klyshko's criterion preceded the WSBY criterion and was
the first proposal for a direct test of nonclassicality according
to `local properties' of the photon number distribution
(PND)~\cite{Kly96},
\begin{equation}
\label{eq:K}
    K_n=(n+1)\frac{p_{n-1}p_{n+1}}{np_n^2} < 1, \; n=1,2,3,\ldots .
\end{equation}
By the replacement $p\mapsto\wp$, the criterion applies to measured photon statistics
rather than the ideal photon counting distribution.
Lee~\cite{Lee98} refers to such tests of nonclassicality according to local properties of
the PND as Type~II, as opposed to the traditional version
(which Lee calls Type~I)
that employs inequalities on moments (such as the Mandel~$Q$ parameter~\cite{Man82}).

If criterion~(\ref{eq:K}) is satisfied for any~$n$, then the field is necessarily nonclassical.
To compare with $\Gamma$,
consider $n=2$ for Klyshko's criterion applied to measured data:
\begin{equation}
\label{eq:K2}
    K_2=\frac{3}{2}\frac{\wp_1\wp_3}{\wp_2^2} < 1.
\end{equation}
Rearranging the terms in Eq.~(\ref{eq:K2}) yields the nonclassicality criterion
\begin{equation}
    \wp_2>\sqrt{\frac{3}{2}\wp_1\wp_3}.
\end{equation}
An alternative criterion emerges from the nonclassicality criterion~(\ref{eq:GammaWDSBY}),
which yields
\begin{equation}
\label{eq:Gamma}
   \wp_2>\frac{1}{2}\sqrt{\frac{3}{2}}(\wp_1+\wp_3).
\end{equation}
Thus, combining Klyshko's result for~$K_2$ and the WDSBY result
for $\Gamma$, we obtain the general condition, based on measured results
$\{\wp_1,\wp_2,\wp_3\}$ for nonclassicality being
\begin{equation}
    \wp_2>\text{min}\left\{\sqrt{\frac{3}{2}\wp_1\wp_3},\frac{1}{2}\sqrt{\frac{3}{2}}(\wp_1+\wp_3)\right\}.
\label{eq:combined}
\end{equation}
This condition is based on the best case by combining the WDSBY
$\Gamma$ criterion~(\ref{eq:GammaWDSBY}) with Klyshko's criterion
for $n=2$.

Eq.~(\ref{eq:combined}) can be further simplified: the two quantities in this equation
coincide in the symmetric case that $\wp_1=\wp_3$:
then $\wp_2>\sqrt{3/2}\wp_1$. Now suppose that $\wp_3=\kappa\wp_1$.
We can thus rewrite condition~(\ref{eq:combined}) as
\begin{equation}
    \frac{\wp_2}{\wp_2}
        >\frac{3}{2}\text{min}\left\{\sqrt{\kappa},\frac{\kappa+1}{2}\right\}.
\label{eq:combined'}
\end{equation}
Except for $\kappa=0,1$ the first term on the left-hand side of Eq.~(\ref{eq:combined'})
is always smaller than the second term, and the two terms are equal when $\kappa=0,1$.
Thus Klyshko's condition is stronger than the WDSBY criterion, and
another advantage of Klyshko's general criterion over WDSBY's criterion is
that the latter only applies in the measured photon number
distribution localized around~$n=2$ whereas Klyshko's criterion
explores nonclassicality for local regions of the photon number
distribution around any $n\ge 1$.

\begin{figure}
\includegraphics[scale=.4]{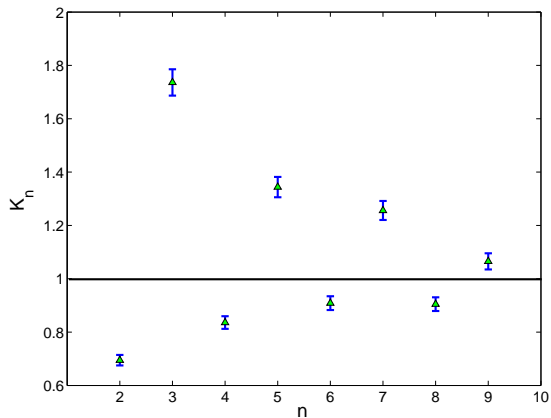}
\caption{\label{fig:KlyshBound} Experimental results for Klyshko's
figure of merit~$K$ vs photon number~$n$ for a single-mode field.}
\end{figure}

In Fig.~\ref{fig:KlyshBound}, we
present a plot of $K_n$ vs~$n$ for the WDSBY data. Klyshko's
criterion is violated for even numbers of photons between $n=2$
and $n=8$. The limit of our measurement capability is nine
photons, and all odd-number detections fail to violate Klyshko's
criterion as expected from joint
distribution~(\ref{eq:2modecounting}). The reason why even-number
counts violate the classical limit and odd-number counts do not is
that the ideal photon number distribution is non-zero only for
even-number counts. Experimentally we detect odd-number counts due
to finite detector quantum efficiency, but this does not lead to
violation of the classical limit.  Fig.~1 is especially important
in demonstrating that nonclassicality of the beam is evident for
all even photon numbers up to the limits imposed by our photon
counter capabilities.

\section{Twin Beam non-classical statistics}

Now let us consider the twin-beam PDC and its joint photon
counting statistics. We will use Lee's generalization of Klyshko's
Type~II criterion to establish nonclassicality of the twin-beam
PDC output directly from the joint photon counting data with no
adjustments made for photodetection efficiency. We will need to
adapt Lee's criterion to accommodate the multi-mode aspect of the
PDC output.

Lee introduces the non-classical criterion for two modes of a
light field given by
\begin{align}
\label{eq:R}
R_{n_1 n_2}=&(n_2+1)\frac{p_{n_1-1,n_2+1}}{2n_1p_{n_1,n_2}} \nonumber \\
    &+(n_1+1)\frac{p_{n_1+1,n_2-1}}{2n_2p_{n_1,n_2}}<1,
\end{align}
with~$n_1,n_2\in\{1,2,3,\ldots\}$.  The field is nonclassical if
$R_{n_1 n_2}$ satisfies inequality~(\ref{eq:R}) for any~$n_1$
and~$n_2$. Although Lee's analysis supposes a two-mode field, this
result is equally valid for a multimode field and the use of two
photon counters, with some modes directed to one detector and
other modes directed to the other detector.

Mode mismatches and lost modes are partially responsible for
photon counter inefficiencies. As we do not employ auxiliary
assumptions, the onus is on us to enhance efficiency to ensure
that inequality~(\ref{eq:R}) is satisfied, and not to resort to
modeling these losses and adjusting the photon statistics
according to these assumptions.

We define $n_1$ as the number of signal photons and $n_2$ as the
number of idler photons.  It is convenient to determine the
conditional probabilities for $n_2$ photons given a count of $n_1$
at the other detector, denoted by~$p_{n_2|n_1}$. The conditional
and joint number counting probabilities are related by the formula
\begin{equation}
    p_{n_1,n_2}=p_{n_2|n_1}p_{n_1}.
\end{equation}
As explained in the previous
section, the counting statistics for each detector is Poissonian
distribution with
\begin{equation}
    \wp_n=\exp(-\bar{n})\bar{n}^n/n!
\end{equation}
where~$\bar{n}$ is the mean photon number summed over all the signal
modes.

The choice of the poisson distribution is justified by the fact
that there are many more signal and idler modes than photons
distributed amongst these modes~\cite{Had05}.
Thus, each mode is predominantly
a vacuum state with a small one-photon contribution and a
negligible multi-photon contribution. The photon distribution
across all modes of each beam is therefore binomial, which, in the
limit of a large number of modes and fixed number of photons, is
Poissonian. This Poissonian distribution contrasts with the
`thermal' photon statistics for an output mode of ideal
nondegenerate PDC, which follows from~(\ref{eq:2modesqvac}).

Lee's expression~(\ref{eq:R}) can be revised in terms of conditional probabilities.
Using the simple relation $\wp^\text{Poisson}_n/\wp^\text{Poisson}_{n-1}=\bar{n}/n$, we rewrite the criterion as
\begin{equation}
\label{eq:Rrevised} R_{n_1
n_2}=\frac{\bar{n}^2\wp_{n_2-1|n_1+1}+n_1(n_1+1)\wp_{n_2+1|n_1-1}}
    {2\bar{n}n_2\wp_{n_2|n_1}}<1.
\end{equation}
In our experiments, the mean photon number collected over all the
modes is $\bar{n}=1$.

\begin{figure}
\includegraphics[scale=.6]{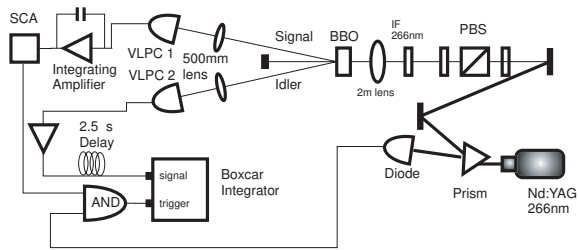}
\caption{\label{fig:setup} Experimental setup for measuring violation of Lee's inequality.}
\end{figure}

The experimental setup for generating twin photon beams for
measuring the Lee non-classical criterion is shown in
Fig.~\ref{fig:setup}. A 266nm pump source is generated from the
fourth harmonic of a Q-switched Nd:YAG laser. The pump pulses have
a duration of 20ns, and a repetition rate of 45kHz.  A dispersion
prism separates the fourth harmonic from the residual second
harmonic, which is used to illuminate a high speed photodiode to
generate a triggering signal.  The fourth harmonic pumps a BBO
crystal, which is set for non-collinear degenerate phase-matching.

In this condition, the signal and idler waves are both 532nm in
wavelength and have a divergence angle of 1 degree from the pump.
The pump is loosely focused before the BBO crystal to achieve a
minimum waist at the collection lens.  This results in a sharper
two-photon image which enhances the collection efficiency.  The
pump power is set to $20\mu W$.  Using the count rate on the
detectors and known values of the detector quantum efficiency, the
average pair creation rate at this pump power is measured to be
one pair per pump laser pulse.

Two VLPC detectors are used in this experiment.  Each detector is
held in a separate helium bath cryostat and cooled down to 6-7K,
which is the optimum operating temperature.  The VLPC is sensitive
to photons with wavelengths of up to 30$\mu$m, so it must be
shielded from room temperature thermal radiation.  This is
achieved by encasing the detector in a copper shield, which is
cooled down to 6K.  Acrylic windows at the front of the copper
shield are used as infra red filters.  These windows are highly
transparent at visible wavelengths and simultaneously nearly
opaque at 2-30$\mu$m wavelengths.

VLPC~1 is used as the triggering detector which detects the number
of photons generated in the signal arm on a given laser pulse. The
output of VLPC~1 is amplified by an integrating amplifier, which
generates an electrical pulse whose height is proportional to the
number of emitted electrons.  The height of the pulse is
discriminated by a single channel analyzer (SCA).  A logical AND
is performed between the output of the SCA and the output of the
photodiode to reject all detection events which occur outside of
the pulse duration.

The output of VLPC~2, which measures photons
in the idler arm, is amplified and connected to the signal input
of a boxcar integrator.  On each pulse from the SCA, the output of
VLPC~2 is integrated over a 2$\mu$s window, which is sufficiently
large to encompass the entire electrical pulse (determined by the
bandwidth of subsequent amplifiers).  The pulse area is
proportional to the number of detected photons.  By measuring the
pulse area histogram of the VLPC, we can therefore measure
$p_{n_2|n_1}$, where $n_1$ and $n_2$ are the number of photons in
the signal and idler arm respectively.


\begin{table}
\caption{\label{tab:LeeBound}Experimental results for Lee's figure
of merit~$R$ vs photon number~$n$ for twin beams}
\begin{ruledtabular}
\begin{tabular}{l|cccr}
     & &  $n_2$ & & \\
    $n_1$ &1 & 2 & 3 & 4\\\hline
    2 & 0.69 $\pm$ 0.023 & 0.27 $\pm$ 0.007 & 0.47 $\pm$ 0.012 & 1.37 $\pm$ 0.04 \\
    3 & 2.23 $\pm$ 0.08 & 0.70 $\pm$ 0.022 & 0.33 $\pm$ 0.008 & 0.47
    $\pm$
    0.012
\end{tabular}
\end{ruledtabular}
\end{table}

The experimental results are given  in Table~\ref{tab:LeeBound}.
Since we can only measure $n_1$ from one to four photons, we only obtain
expressions for the Lee bound for $n_1=2,3$.  In the ideal
case where we have perfect detection efficiency and no dark
counts, we would have $R_{n_1 n_2}=0$ when $n_1=n_2$.  When
$n_1\ne n_2$ the Lee bound would not be well defined because
$p_{n_1,n_2}=0$, causing a divergence.  In the presence of
detection losses, we would still expect the best violations when
$n_1=n_2$, with worse results in the off-diagonal term. As can be
seen in Table~\ref{tab:LeeBound}, this is indeed the case.  When
$n_1=n_2\in\{2,3\}$, we obtain extremely good violations of the Lee
criterion, whereas off-diagonal terms yield worse violations or no
violations at all.

\section{Conclusions}

In summary we have demonstrated directly that twin-beam PDC
produces non-classical light. As PDC is one of the most important
tools for quantum optics, it seems surprising that the
nonclassical nature of PDC output has not been directly verified
before. One reason for not having previously establishing
nonclassicality of twin-beam PDC is the requirement for
sophisticated, modern photon counters and correlations of their
data. Another reason for the novelty of our results is that only
recently has direct testing of nonclassicality via the local
properties of the measured photon count distribution been
understood~\cite{Kly96,Lee98,Wak04}. Our method of directly
measuring photon counts for twin beams applies to multiple-beam
fields and testing the local properties of the measured photon
count distribution provides a valuable, practical means for
establishing nonclassicality of light, especially in cases where
the photons are not anti-bunched.

\emph{Acknowledgments:---} This work was supported in part by the MURI
Center for Photonic Quantum Information Systems (ARO/DTO program
DAAD19-03-1-0199). BCS acknowledges financial support from
iCORE and the Australian Research Council.

\end{document}